\documentclass[12pt]{article}
\usepackage{epsf,amssymb}
\renewcommand\Im{\mathop{\mbox{Im}}}
\renewcommand\Re{\mathop{\mbox{Re}}}

\newcommand{\bea}{\begin{eqnarray}}
\newcommand{\be}{\begin{equation}}
\newcommand{\eea}{\end{eqnarray}}
\newcommand{\ee}{\end{equation}}
\newcommand{\nn}{\nonumber}
\newcommand{\db}{\bar{d}}
\newcommand{\gd}{\gamma_\mu}
\newcommand{\gu}{\gamma^\mu}

\newcommand{\als}{\alpha_s}

\begin{document}

\pagestyle{empty}
\vspace{-0.6in}
\begin{flushright}
DFPD 01/TH/55 \\
FTUV-IFIC-01-1129 \\
RM3-TH/01-16 \\
ROMA-1328/01
\end{flushright}
\vskip 1.2 cm
\centerline{\Large{\bf $B_d$ -- $\bar B_d$ Mixing and the $B_d \to J/
    \psi K_s$ Asymmetry}}
\vskip 0.3 cm
\centerline{\Large{\bf in General SUSY Models}}
\vskip 1.4cm
\centerline{{\bf D.~Be\'cirevi\'c$^{1}$, M.~Ciuchini$^{2}$, E.~Franco$^1$,
V.~Gim\'enez$^3$,  G. Martinelli$^{1}$, }}
\centerline{{\bf  A.~Masiero$^4$, M.~Papinutto$^{5}$, J.~Reyes$^{1}$
and L.~Silvestrini$^1$}}
\vskip 1.cm
\centerline{$^1$ {\sl Dip. di Fisica, Univ. ``La Sapienza''  and INFN,}}
\centerline{{\sl Sezione di Roma, P.le A. Moro 2, I-00185 Rome,
Italy.}}
\centerline{$^2$  {\sl Dip. di Fisica, Univ. di Roma Tre
and INFN, Sezione di Roma III,}}
\centerline{{\sl Via della Vasca Navale 84, I-00146
Roma, Italy}}
\centerline{$^3$ {\sl Dep. de F\'{\i}sica Te\`orica and IFIC, Univ. de
Val\`encia,}}
\centerline{{\sl Dr. Moliner 50, E-46100, Burjassot, Val\`encia,
Spain}}
\centerline{$^4$ {\sl Dip. di Fisica ``G. Galilei'', Univ. di Padova  and INFN,}}
\centerline{{\sl Sezione di Padova, Via Marzolo 8, I-35121 Padua,
Italy.}}
\centerline{$^{5}${\sl Dip. di Fisica, Univ. di Pisa and}}
\centerline{{\sl INFN, Sezione di Pisa, via Buonarroti 2, I-56100 Pisa,
Italy}}
\vskip 1.4cm
\abstract{
We present a next-to-leading order determination of the
gluino-mediated SUSY contributions to $B_d$ -- $\bar B_d$ mixing and
to the CP asymmetry $a_{J/\psi K_s}$ in
the framework of the mass-insertion approximation.
Using hadronic matrix elements recently computed on the lattice,
we obtain improved constraints on the squark-mass splittings.
}
\newpage
\pagestyle{plain}
\setcounter{page}{1}

\section{Introduction}
\label{sec:intro}

With the advent of $B$ factories, $B$ physics is playing a key role in
testing the Standard Model (SM) picture of flavour and CP violation and in probing
 virtual effects from new physics at low energies. In particular, measurements
of $\Delta m_d$, the mass difference in the $B_d$ -- $\bar B_d$ system, and
of $a_{J/\psi K_s}$, the time-dependent CP asymmetry in $B_d \to
J/\psi K_s$ decays, can be used to put stringent constraints on
new-physics contributions to $\Delta B=2$ processes. In terms of the matrix elements of the
effective $\Delta B=2$ Hamiltonian we have
\bea \Delta m_d &=&
2\, \rm{Abs}[\langle\bar  B_d \vert
 {\cal H}_{\rm eff}^{\Delta B=2} \vert B_d \rangle ]
\\ a_{J/\psi K_s} &=&  \sin 2 \beta_{\rm eff} \,  \sin \Delta m_d \, t \, ,\eea
where
\begin{equation}
  \label{eq:asm}
2  \beta_{\rm eff} =
 \, \rm{Arg}[\langle\bar
  B_d \vert {\cal H}_{\rm eff}^{\Delta B=2} \vert B_d \rangle ]\,.
\end{equation}
Within the SM the value of
$\sin 2\beta_{\rm eff}=\sin 2 \beta$ can be connected
to the CP violation angle $\delta$ of the Unitarity Triangle (UT)
\be \sin 2 \beta = \frac{2 \sigma \sin \delta (1-\sigma \cos \delta)}{1 +\sigma^2
-2 \sigma \cos \delta} \, , \ee
where $\sigma = \sqrt{\rho^2 + \eta^2}$ in the Wolfenstein
parametrisation~\cite{Wolfenstein:1983yz}. Assuming that there are no new physics effects  in $B_d$ -- $\bar B_d$ and
$K^0$ -- $\bar K^0$ mixing, 
$\sin 2 \beta$  is  well determined by the standard UT analysis. For
example, ref.~\cite{stocchi} quotes
\begin{equation}
  \label{eq:smbeta}
  \sin 2 \beta =0.698 \pm 0.066.
\end{equation}
The first direct measurements of $a_{J/\psi K_s}$ at the $B$ factories
suggested   a rather low value of the asymmetry~\cite{earlymeas} which, if
confirmed, would have been a hint of new physics
\cite{buras2}--\cite{masierini}. The most recent
world average~\cite{worldaverage}
\begin{equation}
  \label{eq:sin2bexp}
  \sin 2 \beta   = 0.79 \pm 0.10 \, , 
\end{equation}
however, is in very good  agreement with eq.~(\ref{eq:smbeta}) and therefore does not
favour extra contributions. In spite of 
this, it is worth investigating, in  neutral 
$B$ meson systems, signals of  new physics which may emerge in the near
 future thanks to decreasing experimental errors and theoretical uncertainties.
On the one hand, for $\Delta B=2$ transitions, where the SM seems to 
agree with the data,  it is important to improve the theoretical
accuracy   for  constraining  new physics 
contributions  to  box diagrams. On the other,  more accurate 
predictions of  mixing   help to uncover signals of new 
physics in $B$ decay asymmetries, for example $B \to K \pi$ 
or $B \to K \phi$.  These processes are dominated by loop 
contributions and  one  measures  simultaneously the effect of  mixing  and  CP  violation in  decay 
amplitudes, from both SM and new physics virtual particles. Finally, 
of particular interest is the $B_{s}$--$\bar B_{s}$ mixing for which 
the SM amplitude is real and 
the phase  entirely originates from new physics effects. This may
give rise to CP violating effects which would be absent in the SM. 

In this work, we aim at testing new-physics effects in $\Delta B=2$
transitions using the most recent experimental and theoretical
results: the new data from $B$ factories and the very recent lattice
determinations of the relevant matrix elements, which allow the 
computation of $\Delta m_d$ and of $a_{J/\psi K_s}$ at the
Next-to-Leading Order (NLO), apart from some additional  model-dependent 
${\cal O} (\alpha_s(M_W))$ terms. First of all, we present
model-independent formulae for $\Delta B=2$ ($B_d$ and $B_s$)
transitions  which include all possible  new-physics contributions. These formulae
are obtained by considering the most general effective Hamiltonian
for $\Delta B=2$ processes, ${\cal H}_{\rm eff}^{\Delta B=2}$.
We write down the expression for $\Delta m_d$ ($\Delta m_s$) and for  
$\beta_{\rm eff}$
in terms of the Wilson coefficients at the electroweak scale, including
NLO QCD corrections in the running from $M_W$ to $\mu=m_b$~\cite{Ciuchini,Urban}
and the matrix elements very recently computed in lattice
QCD~\cite{Damir}, in the same RI-MOM renormalisation scheme. These expressions can be readily
used to compute $\Delta m_d$
and $\beta_{\rm eff}$ in any extension of the SM.

As a concrete example of physics beyond the SM, we then consider the  
Minimal Supersymmetric Standard Model 
(MSSM) with arbitrary soft breaking terms and we study model-independent constraints on the
SUSY parameter space coming from $\Delta m_d$ and $a_{J/\psi K_s}$. We
closely follow the analysis of ref.~\cite{noiK}, where upper bounds
on new sources of flavour and CP violation were obtained from the study of $K^0$--$\bar K^0$
mixing. For earlier analyses, either at tree level or with LO evolution, using  na\"{\i}ve
Vacuum Insertion Approximation (VIA) $B$ parameters,
see refs.~\cite{antichi} and \cite{bagger} respectively.
Ref.~\cite{noiK} and the present work can be considered as  first
steps toward a full NLO model-independent analysis of
flavour and CP violation beyond the SM, with special focus on its
most studied extension, namely the general MSSM. We stress that the
inclusion of hadronic matrix elements from lattice QCD,
for operators renormalised consistently with the Wilson
coefficients at the NLO, is highly important to  reduce the uncertainties in the study of flavour and
CP violation. Indeed, while  matrix elements for
 operators appearing in the SM have been studied for a long
time, it was only recently that similar results have been obtained for
operators that only come from physics  beyond the 
SM~\cite{Damir,allton}. More work
 is still to be done along these lines.  For example, the large uncertainties
in  the SUSY constraints  from the electric dipole moment of the
neutron  could be drastically reduced by a lattice computation
of the relevant matrix elements~\cite{guidoberlin}.

The paper is organised as follows. In the next section we introduce the
most general ${\cal H}_{\rm eff}^{\Delta B=2}$ with the inclusion of
the NLO QCD corrections in the evolution from the scale of new physics
down to low energy. We  provide the expression of the Wilson
coefficients at the low  hadronic scale ($\sim m_{b}$) as a
function of the Wilson coefficients and of $\alpha_s$ at the ``new
physics scale''.  Section \ref{sec:ME} deals with the evaluation of the
hadronic matrix elements of the local operators appearing in  ${\cal H}_{\rm
  eff}^{\Delta B=2}$. Here we  replace the ``traditional'' values
of the $B$ parameters in the VIA with the values  recently
obtained from a lattice computation in the RI-MOM scheme~\cite{Damir}.  We then turn to SUSY
and present in Section \ref{sec:SUSY} the expression of
gluino-mediated contributions to ${\cal H}_{\rm eff}^{\Delta B=2}$. 
As long as we deal with general squark mass matrices, the inclusion of the gluino-mediated FCNC
alone is sufficient to get the correct bulk of the SUSY contribution.
In specific models  corresponding to particularly restricted squark 
mass  matrices, however, there are regions
of the SUSY parameter space where charginos, stops and/or charged Higgs are relatively light
and their exchange gives contributions even larger than  gluinos to FCNC~\cite{tre}.
This  happens more easily  in the case of $B_d$ -- $\bar B_d$ mixing, rather than for  $K^0$--$\bar K^0$,
because the relevant mixing angles and/or Yukawa couplings are larger in this case.
Unfortunately, the experimental information and the theoretical accuracy are 
presently insufficient to attempt
a simultaneous analysis  which includes a spanning of the squark mass matrices and of  the
stop/chargino mass values. For this reason, in this first study which uses the new
measurements of $\sin 2 \beta$, NLO corrections and lattice $B$ parameters, we
include gluino-mediated FCNC only.  
Our quantitative results are presented in Sect.~\ref{sec:numerics} and
summarised in  Tables~\ref{tab:reds2_500} and \ref{tab:reds2_1250}.
We  perform our computations in the mass insertion
approximation~\cite{hall}.  We choose the super-CKM basis for the
fermion and sfermion states, where all the couplings of these particles
to neutral gauginos are flavour diagonal, while  genuine SUSY FC
effects are exhibited by the non-diagonal entries of the sfermion mass
matrices. Denoting by $\Delta^2$ the off-diagonal terms in the sfermion
mass matrices (i.e.  the mass terms relating sfermions of the same
electric charge, but different flavour), the sfermion propagators can
be expanded as a series in terms of $\delta = \Delta^2/ \tilde{m}^2$,
where $\tilde{m}$ is the average sfermion mass.  As long as $\Delta^2$
is significantly smaller than $\tilde{m}^2$, we can just take the
first term of this expansion and then the experimental information
concerning FCNC and CP violating phenomena translates into upper
bounds on the $\delta$s~\cite{noiK,antichi}.
Tables~\ref{tab:reds2_500}  and \ref{tab:reds2_1250} contain new
constraints on the  parameters $\delta$.
With respect to previous analyses, we find that the inclusion  of the direct measurements
of $\sin 2 \beta$  allows imposing  quite stringent constraints on $\Im \delta_{13}$~\cite{Chua:2001dd}.
On the theoretical side, as already observed in~\cite{noiK}, the  QCD evolution from large scales
and the use of the lattice $B$ parameters  induces sizable changes in the limits on the
values of the $\delta$s. The major effect of our improved computation is felt
by Left-Right operators.

\section{Effective Hamiltonian for $\Delta B=2$ processes beyond the SM}
\label{sec:EH}

The most general effective Hamiltonian for $\Delta B=2$ processes
beyond the SM has the following form:
\begin{equation}
  \label{eq:defheff}
  {\cal H}_{\rm eff}^{\Delta B=2}=\sum_{i=1}^{5} C_i\, Q_i +
  \sum_{i=1}^{3} \tilde{C}_i\, \tilde{Q}_i
\end{equation}
where
\begin{eqnarray}
         \label{Qi}
        Q_1 & = & \db^{\alpha}_L \gd b^{\alpha}_L \db^{\beta}_{L} \gu
        b^{\beta}_L\; ,
        \nonumber \\
        Q_2 & = & \db^{\alpha}_R  b^{\alpha}_L \db^{\beta}_R b^{\beta}_L\; ,
        \nonumber \\
        Q_3 & = & \db^{\alpha}_R  b^{\beta}_L \db^{\beta}_R b^{\alpha}_L\; ,
       \\
        Q_4 & = & \db^{\alpha}_R  b^{\alpha}_L \db^{\beta}_L b^{\beta}_R\; ,
        \nonumber \\
        Q_5 & = & \db^{\alpha}_R  b^{\beta}_L \db^{\beta}_L b^{\alpha}_R\; ,
       \nonumber
\end{eqnarray}
and the operators $\tilde{Q}_{1,2,3}$ are obtained from the
$Q_{1,2,3}$ by the exchange $ L \leftrightarrow R$.  Here
$q_{R,L}=P_{R,L}\,q$, with $P_{R,L}=(1 \pm \gamma_5)/2$, and $\alpha$
and $\beta$ are colour indexes. In the case of the $B_s$ system  one has simply to replace the $d$ with the $s$
quark in the operators appearing in eqs.~(\ref{Qi}).

The NLO anomalous dimension matrix for the most general
${\cal H}_{\rm eff}^{\Delta F=2}$ has been computed in~\cite{Ciuchini}
and the results have been confirmed in~\cite{Urban}.  We use the
Regularisation-Independent anomalous dimension matrix in the Landau gauge
(also known as RI-MOM), since we will make use of matrix elements computed in lattice
QCD with the same choice of renormalisation scheme.

A full NLO computation would also require the $O(\alpha_s)$
corrections to the matching conditions which determine the Wilson 
coefficients, see eqs.~(\ref{inicoeff}) below.  These
corrections are model-dependent and have been computed only in few specific 
cases~\cite{calcu,feng}.  One
might argue that, being of order $\alpha_s(M_S)$ (here and in
the following $M_S$ represents the scale of new physics), these
contributions should be small, as suggested by the cases of the SM and
of the two Higgs doublet model~\cite{calcu}. This statement, however,  can only be
confirmed by explicit computations in specific models.  Due to the
absence of $O(\alpha_s)$ corrections to the matching, our ${\cal
  H}_{\rm eff}^{\Delta F=2}$ are  affected by a residual scheme
dependence, which would be cancelled by the terms of order
$\alpha_s(M_S)$ in the Wilson coefficients $C_i(M_S)$. These terms are 
model dependent. 

\begin{table}[t]
\begin{center}
\begin{tabular}{|c|c|c|c|}
        \hline \hline
        Constant & Value  & Constant & Value  \\[0.2cm]
        \hline
        \hline
        $m_{B_{d}}$ & $5.279$ GeV  &
%
        $f_{B_{d}}$ & $200\pm 30$~MeV  \\[0.2cm]
        $m^{\overline{MS}}_{d}(2 {\rm GeV})$ & $7$ MeV   &
        $m^{\overline{MS}}_{b}(m^{\overline{MS}}_b)$ & $4.23$ GeV  \\[0.2cm]
        $m_{t}
        $ & $174 \pm 5$ GeV  &
        $\als(M_{Z})$ & $0.119$ \\[0.2cm]
        $\vert V_{cb}\vert$  & $(40.7 \pm 1.9 ) \times 10^{-3} $    &  
        $\vert V_{ub}\vert$ &  $(3.61 \pm 0.46 ) \times 10^{-3} $   \\[0.2cm]
        \hline \hline
\end{tabular}
\end{center}
\caption{{\sl Input quantities used in the phenomenological analysis. 
$m^{\overline{MS}}_{d}$ and $m^{\overline{MS}}_{b}$ are the 
$\overline{MS}$ masses whereas $m_{t}$ is the pole mass.}}
\label{tab:parameters}
\end{table}

The  $C_i(M_S)$ are obtained by integrating out all
new particles simultaneously at the scale $M_S$. We then have to
evolve the coefficients down to the hadronic scale 
$\mu=m_b=4.6$~GeV ($m_{b}\equiv m_{b}(\mu= m_{b})$ is the RI-MOM mass),
which is the renormalisation scale of the
operators used in ref.~\cite{Damir}.  By coincidence,  $m_{b}$ 
has the same numerical value as the pole mass,
$m_b^{pole}=4.6$~GeV,   extracted at the NLO  from the $\overline{MS}$
mass $m^{\overline{MS}}_{b}(m^{\overline{MS}}_{b})=4.23$~GeV, see 
Table~\ref{tab:parameters}. 
For  consistency, we have to evolve the Wilson coefficients at the same  
renormalisation scale.  The SM contribution can be computed independently and evolved from $M_W$ to $\mu$ using the
well-known NLO QCD corrections~\cite{burasold}.

We give here an analytic formula for the expression of the Wilson
coefficients at the scale $\mu=m_b$   as a function of the initial
conditions at the SUSY scale $C_i(M_S)$ and of $\alpha_s(M_S)$. This
formula has been obtained by using, for the SM parameters, the values in
Table~\ref{tab:parameters}.
For $M_S>m_t$ we obtain
\begin{equation}
\label{eq:magic1}
C_r(m_b^{pole})=\sum_i \sum_s
             \left(b^{(r,s)}_i + \eta \,c^{(r,s)}_i\right)
             \eta^{a_i} \,C_s(M_S),
\end{equation}
where, in the evolution of the coefficients from $M_S$, we have chosen
$M_S= (M_{\tilde g} + M_{\tilde q})/2$ and 
$\eta=\alpha_s(M_S)/\alpha_s(m_t)$.  The magic numbers
are given below: 
\begin{equation}
\label{eq:magic2}
\begin{array}{l l}
a_i=(0.286, -0.692, 0.787, -1.143, 0.143)& \\
& \\
b^{(11)}_i=(0.865, 0, 0, 0, 0),&
c^{(11)}_i=(-0.017,0,0,0,0),\\
b^{(22)}_i=(0,1.879,0.012,0,0),&
c^{(22)}_i=(0,-0.18,-0.003,0,0),\\
b^{(23)}_i=(0,-0.493,0.18,0,0),&
c^{(23)}_i=(0,-0.014,0.008,0,0),\\
b^{(32)}_i=(0,-0.044,0.035,0,0),&
c^{(32)}_i=(0,0.005,-0.012,0,0),\\
b^{(33)}_i=(0,0.011,0.54,0,0),&
c^{(33)}_i=(0,0.000,0.028,0,0),\\
b^{(44)}_i=(0,0,0,2.87,0),&
c^{(44)}_i=(0,0,0,-0.48,0.005),\\
b^{(45)}_i=(0,0,0,0.961,-0.22),&
c^{(45)}_i=(0,0,0,-0.25,-0.006),\\
b^{(54)}_i=(0,0,0,0.09,0),&
c^{(54)}_i=(0,0,0,-0.013,-0.016),\\
b^{(55)}_i=(0,0,0,0.029,0.863),&
c^{(55)}_i=(0,0,0,-0.007,0.019),\\
\end{array}
\end{equation}
and we have only written the non-vanishing entries. The magic numbers
for the evolution of $\tilde C_{1-3}$ are the same as the ones for the
evolution of $C_{1-3}$.  Formulae~(\ref{eq:magic1}) and
(\ref{eq:magic2}) are combined with the $B$ parameters
evaluated at $m_b^{pole}=4.6$~GeV in the MOM-RI scheme~\cite{Damir},
given in eqs.~(\ref{eq:fres}) below~\footnote{Magic numbers in the NDR scheme
can be found in ref.~\cite{Buras:2001ra}.}.
In this way it is possible to  compute the
contribution to $\Delta m_{d,s}$ and $\beta_{\rm eff}$ at the NLO in QCD for
any model of new physics in which the new contributions with respect
to the SM originate from extra heavy particles. One just has to plug
in the expression for the $C_i$ evaluated at the large energy scale
$M_S$ in his favourite model. When the $O(\alpha_s)$ corrections to
the $C_i(M_S)$ are available, one can obtain a full NLO,
regularisation-independent result; in the cases where this corrections
have not been computed yet, the results contain a residual systematic
uncertainty of order $\alpha_s(M_S)$. We note that, due to the
presence of large entries in the NLO anomalous dimension matrix, a
systematic uncertainty is present in the QCD
evolution from the SUSY scale to the hadronic one.  This  uncertainty
stems from different ways of writing the NLO evolution matrix, which 
differ by terms of ${\cal O}(\alpha^{2}_{s})$,  and  has been taken into 
account in the numerical analysis as  uncertainty in the numerical 
value of the magic numbers.  For this reason  we have 
presented only the significant figures of the magic numbers, which 
have an uncertainty of one unit on the last digit. The only exception 
is   $c^{(44)}_4$ for which the uncertainty is 3 units.

\section{Hadronic Matrix Elements}
\label{sec:ME}

The matrix elements of the operators $Q_i$ between neutral $B$ mesons in the
VIA are given by:
\begin{eqnarray}
        \label{me}
        \langle \bar B_d \vert Q_{1}
        \vert B_d \rangle_{\rm VIA} & = &
        \frac{1}{3} m_{B_d} f_{B_d}^{2}\; ,
        \nonumber \\
        \langle \bar B_d  \vert Q_{2}
        \vert B_d \rangle_{\rm VIA}  & = & -\frac{5}{24}
        \left(\frac{m_{B_d}}{m_{b}+m_{d}}\right)^{2}  m_{B_d} f_{B_d}^{2}\; ,
        \nonumber \\
        \langle \bar B_d \vert Q_{3}
        \vert B_d \rangle_{\rm VIA}  & = & \frac{1}{24}
        \left(\frac{m_{B_d}}{m_b{}+m_{d}}\right)^{2} m_{B_d} f_{B_d}^{2}\; ,
        \\
        \langle \bar B_d  \vert Q_{4}
        \vert B_d \rangle_{\rm VIA} & = & \left[\frac{1}{24} +
        \frac{1}{4}
        \left(\frac{m_{B_d}}{m_{b}+m_{d}}\right)^{2}\right] m_{B_d} f_{B_d}^{2}\; ,
        \nonumber \\
        \langle \bar B_d  \vert Q_{5}
        \vert B_d \rangle_{\rm VIA} & = &\left[\frac{1}{8} +
        \frac{1}{12}
        \left(\frac{m_{B_d}}{m_{b}+m_{d}}\right)^{2}\right]  m_{B_d} f_{B_d}^{2} \; ,
        \nonumber
\end{eqnarray}
where $m_{B_d}$ is the mass of the $B_d$ meson and $m_{b}$ and $m_{d}$ are
the  masses of the $b$ and $d$ quarks respectively.    Note that the
normalisation of the operators used in this paper differs by a factor
of $2  m_{B_d}$ from the one used in ref.~\cite{Damir}. Here and in the
following, the same expressions of the $B$ parameters of the operators
$Q_{1-3}$ are valid for the operators $\tilde Q_{1-3}$, since strong
interactions preserve parity.  In the case of the $B_s$ system  one has simply to replace the $d$ with the $s$
quark in the operators and expressions appearing in eqs.~(\ref{me}).

In the case of the renormalised operators, we define the
$B$ parameters as follows:
\bea
\label{eq:bpars}
\langle \bar B_d \vert \hat Q_{1} (\mu)
\vert B_d \rangle & = &
\frac{1}{3} m_{B_d} f_{B_d}^{2}  B_1(\mu)\; ,
\nonumber \\
\langle
\bar B_d \vert \hat Q_{2} (\mu) \vert B_d \rangle &=& -\frac{5}{24}
\left( \frac{ m_{B_d} }{ m_{b}(\mu) + m_d(\mu) }\right)^{2}
 m_{B_d} f_{B_d}^{2}  B_{2}(\mu) \nonumber  \\
\langle \bar B_d \vert \hat Q_{3} (\mu) \vert B_d \rangle &=&
\frac{1}{24} \left( \frac{ m_{B_d} }{ m_{b}(\mu) + m_d(\mu) }\right)^{2}
 m_{B_d} f_{B_d}^{2}  B_{3}(\mu) \\
\langle \bar B_d \vert \hat Q_{4} (\mu) \vert B_d\rangle &=& \frac{1}{4}
\left( \frac{ m_{B_d} }{ m_{b}(\mu) + m_d(\mu) }\right)^{2}
 m_{B_d} f_{B_d}^{2} B_{4}(\mu) \nn \\
\langle \bar B_d \vert \hat Q_{5} (\mu) \vert B_d \rangle &=&
\frac{1}{12} \left( \frac{m_{B_d}  }{ m_{b}(\mu) + m_d(\mu) }\right)^{2}
 m_{B_d} f_{B_d}^{2}  B_{5}(\mu) \nn\ ,
\eea
where the notation $\hat Q_{i}(\mu)$ (or simply $\hat Q_{i}$) denotes
the operators renormalised at the scale $\mu$.  For consistency with the calculation of the $B$ parameters
in ref.~\cite{Damir},   the quark masses, evolved at the scale $\mu$,  must   be computed
in the Landau RI-MOM scheme~\cite{NPM,bibbia}.
With the numbers in Table~\ref{tab:parameters},
this corresponds to $m_{b}=m_b(m_b)= 4.6$~GeV and $m_d(m_b)= 5.4$~MeV  (RI-MOM)~\footnote{
The precise value of $m_d$ is not very important for our analysis.}.

A few words of explanation are necessary at this point. The
$B$ parameter of the matrix element $\langle \bar B_d \vert Q_1 \vert
B_d \rangle$  has been extensively studied on
the lattice due to its phenomenological relevance~\cite{bernard2000,ryan2001},
and used in many UT analyses~\cite{stocchi,hocker}. For the other
operators, instead, all the studies beyond the SM
have used the VIA $B$ parameters, which in some cases, as will be
shown below, is a  crude approximation. The numerical results for the 
$B$ parameters
$B_{i}(\mu)$  refer to the  RI-MOM scheme.
We have used  the same definition of the $B$ parameters
as in ref.~\cite{Damir} from which we have taken the values.
In our numerical study, we have used
\be
\label{eq:fres}
\begin{tabular}{ll}
$B_{1}(m_b) =0.87(4)^{+5}_{-4}$ \, ,   & $B_{2}(m_b) = 0.82(3)(4)$\, ,  \\
$B_{3}(m_b)=1.02(6)(9)$\, ,  &  $B_{4}(m_b) = 1.16(3)^{+5}_{-7}$\, , \\
$B_{5}(m_b) = 1.91(4)^{+22}_{-7}$\, .  & \\
\end{tabular}
\ee
We added a conservative $\pm 10$\% systematic
error to the errors quoted above,
based on previous experience on discretisation, extrapolation  and quenching effects.
In the  cases where a comparison
has been possible, the quenching effects for all the 
$B$ parameters computed so far have always been found much smaller than all 
other uncertainties.

\section{Effective Hamiltonian for $\Delta B=2$ processes in SUSY}
\label{sec:SUSY}

In this Section, we describe the computation of gluino-mediated
contributions to ${\cal H}_{\rm eff}^{\Delta B=2}$ at the NLO in QCD.

Several cases may be considered: i) $m_{\tilde{q}} \sim
m_{\tilde{g}}$, ii) $m_{\tilde{q}} \ll m_{\tilde{g}}$ and iii)
$m_{\tilde{q}} \gg m_{\tilde{g}}$ , where $m_{\tilde{q}}$ is the
average squark mass and $m_{\tilde{g}}$ is the gluino mass.
Case ii), in which $m_{\tilde{q}} \ll m_{\tilde{g}}$, cannot be
realised in the framework considered here, in which the soft SUSY
breaking terms are introduced at the Planck scale. This is due to the
fact that the evolution from the Planck to the electroweak scale
forbids such a mass hierarchy. In fact, neglecting the effects of
Yukawa couplings and A-terms, one obtains in the down sector the
following approximate expression for the ratio
$x=m^2_{\tilde{g}}/m_{\tilde{q}}^2$ at the electroweak scale, in terms
of the value $x_0$ at the Planck scale~\cite{louis}:
\begin{equation}
  \label{xlimit}
  x\simeq \frac{9 x_0}{1+7 x_0} \longrightarrow \frac{9}{7}\, .
\end{equation}
Even if one starts at the super-large scale with an extreme hierarchy
between squark and gluino masses ($x_0 \gg 1$), at the electroweak
scale the two masses will be of the same order.  For this reason, we
will not consider the case $m_{\tilde{q}} \ll m_{\tilde{g}}$ in our
analysis.  Case iii), in which $m_{\tilde{q}} \gg m_{\tilde{g}}$, can
be realised in some special models, such as effective supersymmetry~\cite{kaplan}
or models with a light gluino~\cite{farrar}. The NLO corrections in this case have been computed in
ref.~\cite{ignazio}.  In this first study, however, we do not consider these cases and  present only results for
the case  $m_{\tilde{q}} \sim m_{\tilde{g}}$.

The mass insertion approximation,  which we have adopted in our analysis,  presents the major advantage
that one does not need the full diagonalisation of the sfermion mass
matrices to perform a test of FCNC in  the SUSY model under consideration. 
It is enough to compute ratios of the off-diagonal
over the diagonal entries of the sfermion mass matrices and compare
the results with the general bounds on the $\delta$s which we provide
here from $\Delta m_{B_d}$ and $a_{J/\psi K_s}$. This formulation of the
mass insertion approximation in terms of the parameters $\delta$ is
particularly suitable for model-independent analyses, but involves two
further assumptions: the smallness of the off-diagonal mass terms with
respect to the diagonal ones, and the degeneracy of the diagonal mass
terms in the super-CKM basis. The latter assumption, related to the
use of the average squark mass $\tilde{m}$, is well justified in our
case, since, on quite general grounds, one does not expect a sizable
non-degeneracy of the down-type squarks.
In SUSY GUT it may happen that the sbottom mass is smaller 
than the $d$ and $s$ squark masses. Also in this case, namely 
for non-degenerate diagonal mass terms,  however, 
it is possible to define a generalised mass insertion
approximation~\cite{romanino} and apply the methods exposed in this paper.

\begin{figure}[t]
\begin{center}
\epsfysize=0.4\textheight
\epsfxsize=0.8\textwidth
\epsffile{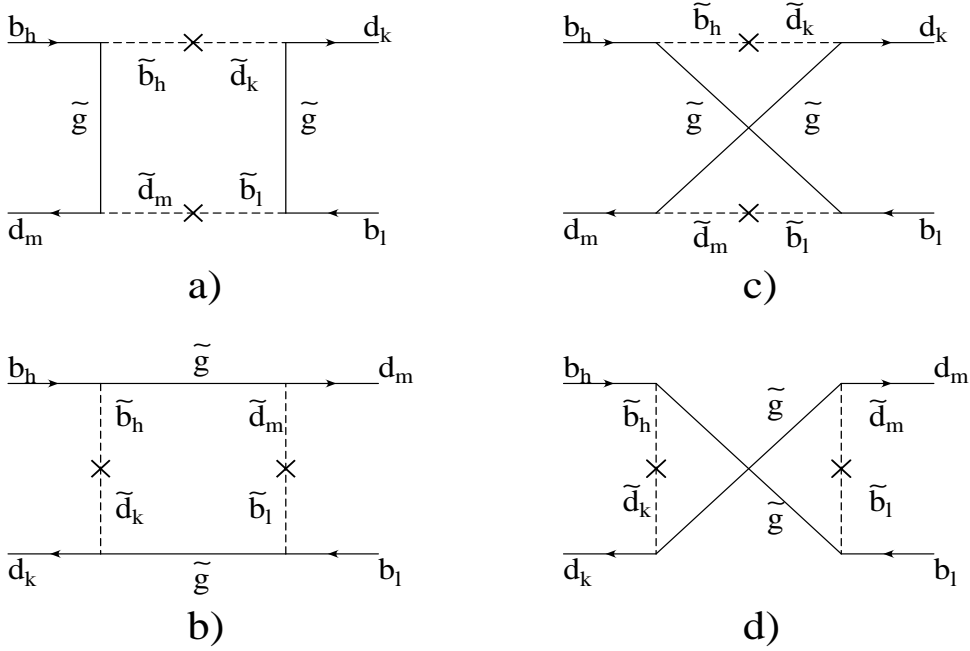}
\end{center}
\caption{{\sl Feynman diagrams for $\Delta B=2$ transitions, with
  $h,k,l,m=\{L,R\}$.}}
\label{fig:ds2}
\end{figure}

There exist four different $\delta$ mass-insertions connecting
flavours $d$ and $b$ along a sfermion propagator:
$\left(\delta^d_{13}\right)_{LL}$, $\left(\delta^d_{13}\right)_{RR}$,
$\left(\delta^d_{13}\right)_{LR}$ and
$\left(\delta^d_{13}\right)_{RL}$.  The indexes $L$ and $R$ refer to
the helicity of the fermion partners.  The amplitude for $\Delta B=2$
transitions in the full theory at the LO is given by the computation
of the diagrams in fig.~\ref{fig:ds2}.
At the lowest order in QCD, in the basis  of
  eq.~(\ref{Qi}),  the Wilson coefficients are given
by~\cite{antichi}:
\begin{eqnarray}
        \label{inicoeff}
        C_1&=&-\frac{\als^2}{216 m_{\tilde{q}}^2}
        \left(  24\,x\,f_6(x) + 66\,\tilde{f}_6(x) \right)
        \left(\delta^d_{13}\right)^2_{LL} \,,
        \nonumber \\
        C_2&=&-\frac{\als^2}{216 m_{\tilde{q}}^2}
         \, 204\,x\,f_6(x) \left(\delta^d_{13}\right)^2_{RL} \,,
        \nonumber \\
        C_3&=& \frac{\als^2}{216 m_{\tilde{q}}^2}
        \,36\,x\,f_6(x) \left(\delta^d_{13}\right)^2_{RL} \,,
        \nonumber \\
        C_4&=&-\frac{\als^2}{216 m_{\tilde{q}}^2}
        \biggl[  \left( 504\,x\,f_6(x) - 72\,\tilde{f}_6(x)\right)
        \left(\delta^d_{13}\right)_{LL}\left(\delta^d_{13}\right)_{RR}
        \nonumber \\
        &&- 132\,\tilde{f}_6(x)
        \left(\delta^d_{13}\right)_{LR}\left(\delta^d_{13}\right)_{RL}
        \biggr] \,,
        \nonumber \\
        C_5&=&-\frac{\als^2}{216 m_{\tilde{q}}^2}
        \biggl[  \left( 24\,x\,f_6(x) + 120\,\tilde{f}_6(x)\right)
        \left(\delta^d_{13}\right)_{LL}\left(\delta^d_{13}\right)_{RR}
        \nonumber \\
        &&- 180\,\tilde{f}_6(x)
        \left(\delta^d_{13}\right)_{LR}\left(\delta^d_{13}\right)_{RL}
        \biggr] \,,
        \nonumber \\
        \tilde{C}_1&=& -\frac{\als^2}{216 m_{\tilde{q}}^2}
        \left(  24\,x\,f_6(x) + 66\,\tilde{f}_6(x) \right)
        \left(\delta^d_{13}\right)^2_{RR} \,,
        \nonumber \\
        \tilde{C}_2&=&-\frac{\als^2}{216 m_{\tilde{q}}^2}
         \, 204\,x\,f_6(x) \left(\delta^d_{13}\right)^2_{LR} \,,
        \nonumber \\
        \tilde{C}_3&=& \frac{\als^2}{216 m_{\tilde{q}}^2}
        \,36\,x\,f_6(x) \left(\delta^d_{13}\right)^2_{LR} \,,
\end{eqnarray}
where $x=m^2_{\tilde{g}}/m_{\tilde{q}}^2$ and the functions $f_6(x)$ and
$\tilde{f}_6(x)$ are given by:
\begin{eqnarray}
f_6(x)&=&\frac{6(1+3x)\ln x +x^3-9x^2-9x+17}{6(x-1)^5}\; , \nonumber  \\
\tilde{f}_6(x)&=&\frac{6x(1+x)\ln x -x^3-9x^2+9x+1}{3(x-1)^5}\; .
\end{eqnarray}
In the absence of   $O(\alpha_s)$ corrections to the matching, we
interpret the $C_i$ given above as coefficients computed at the large
energy scale $M_S\sim m_{\tilde q} \sim m_{\tilde g}$, i.e. $C_i
\equiv C_i(M_S)$.

\begin{figure}[t]
\begin{center}
\begin{tabular}{c c}
\epsfxsize=0.45\textwidth\epsffile{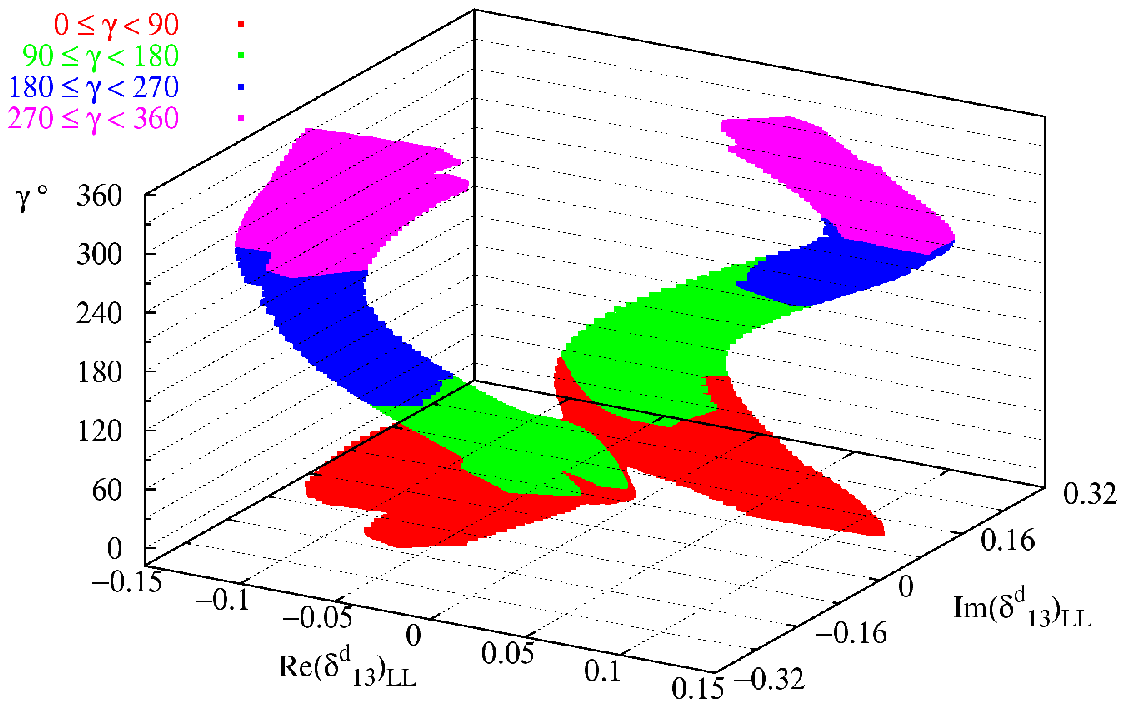} & \epsfxsize=0.45\textwidth\epsffile{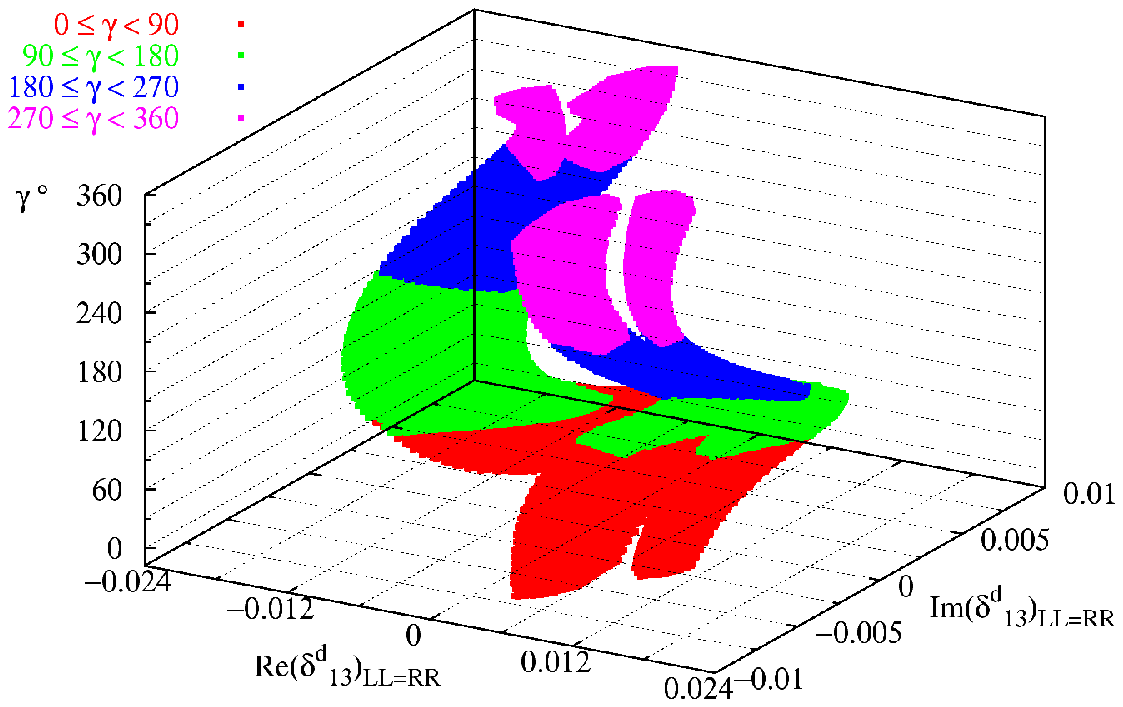}\\
\epsfxsize=0.45\textwidth\epsffile{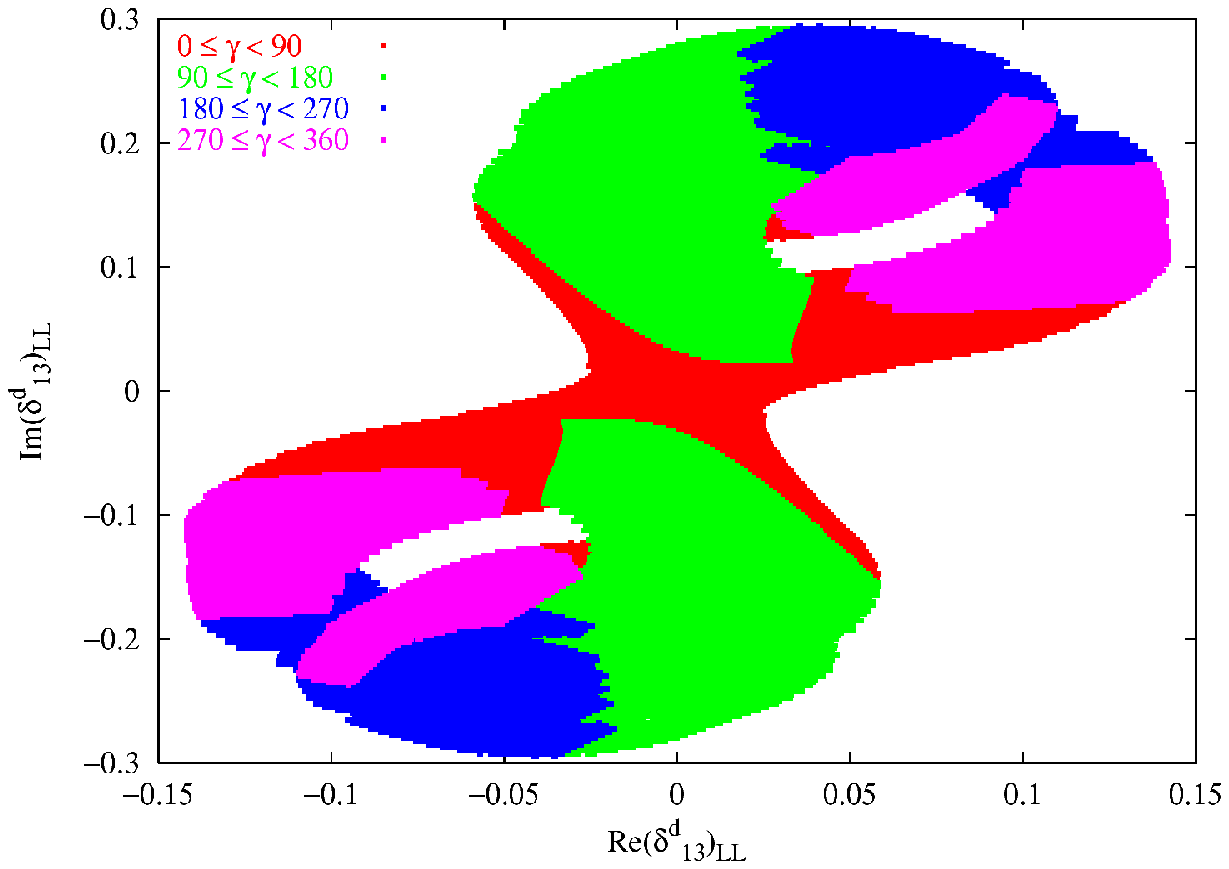} &\epsfxsize=0.45\textwidth \epsffile{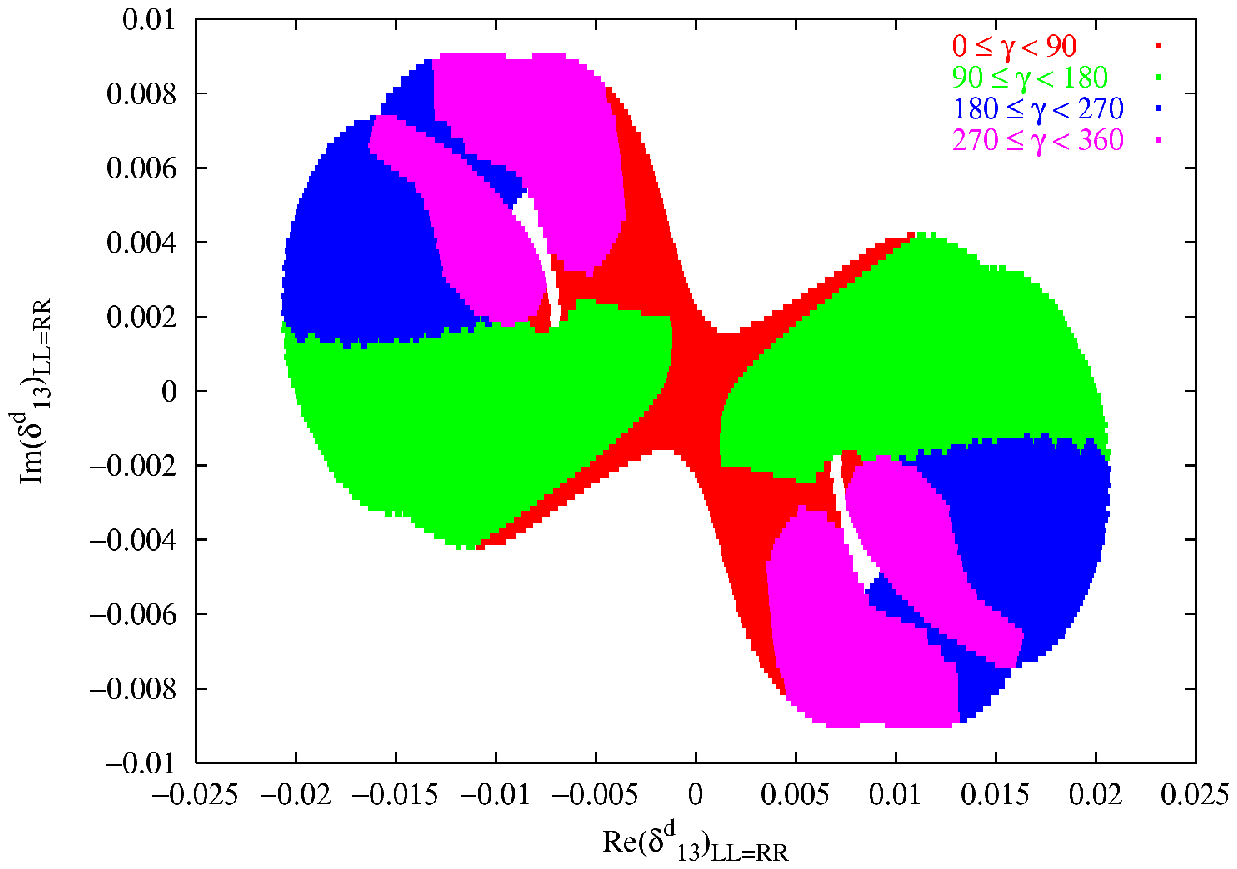}\\
\end{tabular}
\end{center}
\caption{{\sl Allowed regions in the ($\gamma$,  $\Re(\delta^{d}_{13})_{LL}$, $\Im(\delta^{d}_{13})_{LL}$) space
with $(\delta^{d}_{13})_{LL}$ only (left) and
$(\delta^{d}_{13})_{LL}=(\delta^{d}_{13})_{RR}$
(right). The two lower plots are the corresponding projections in the
$\Re(\delta^{d}_{13})_{LL}$--$\Im(\delta^{d}_{13})_{LL}$ plane. Different colours
denote values of $\gamma$ belonging to different quadrants.}}
\label{fig:LL}
\end{figure}

\begin{figure}[t]
\begin{center}
\begin{tabular}{c c}
\epsfxsize=0.45\textwidth\epsffile{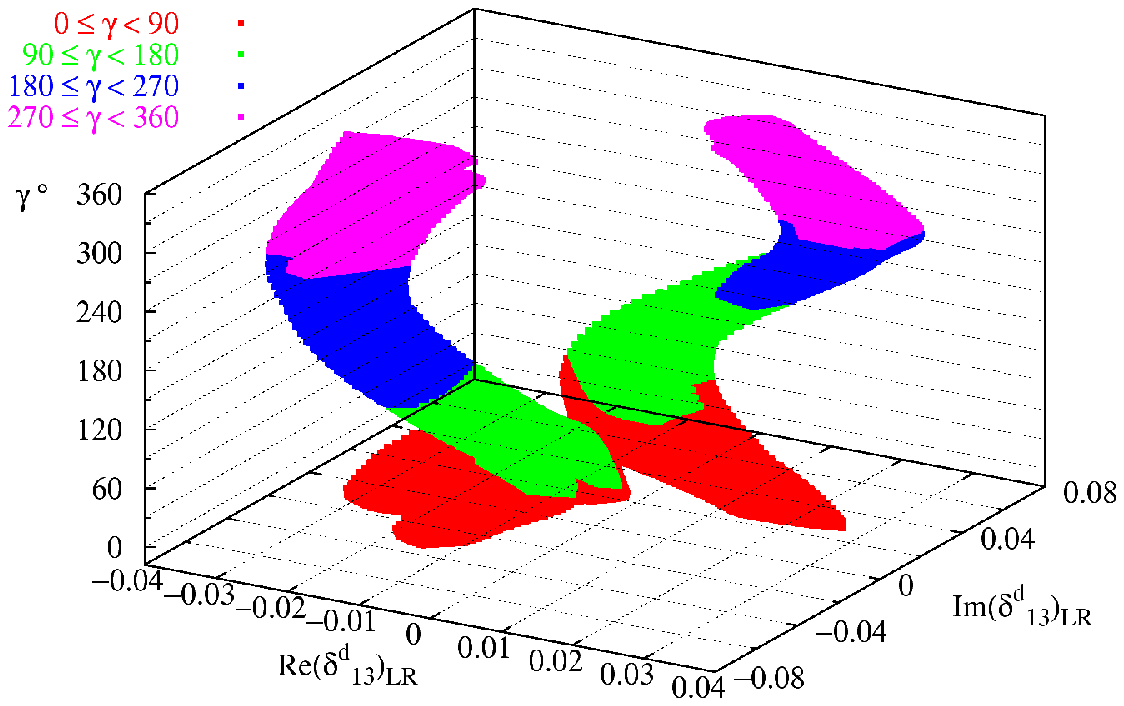} & \epsfxsize=0.45\textwidth\epsffile{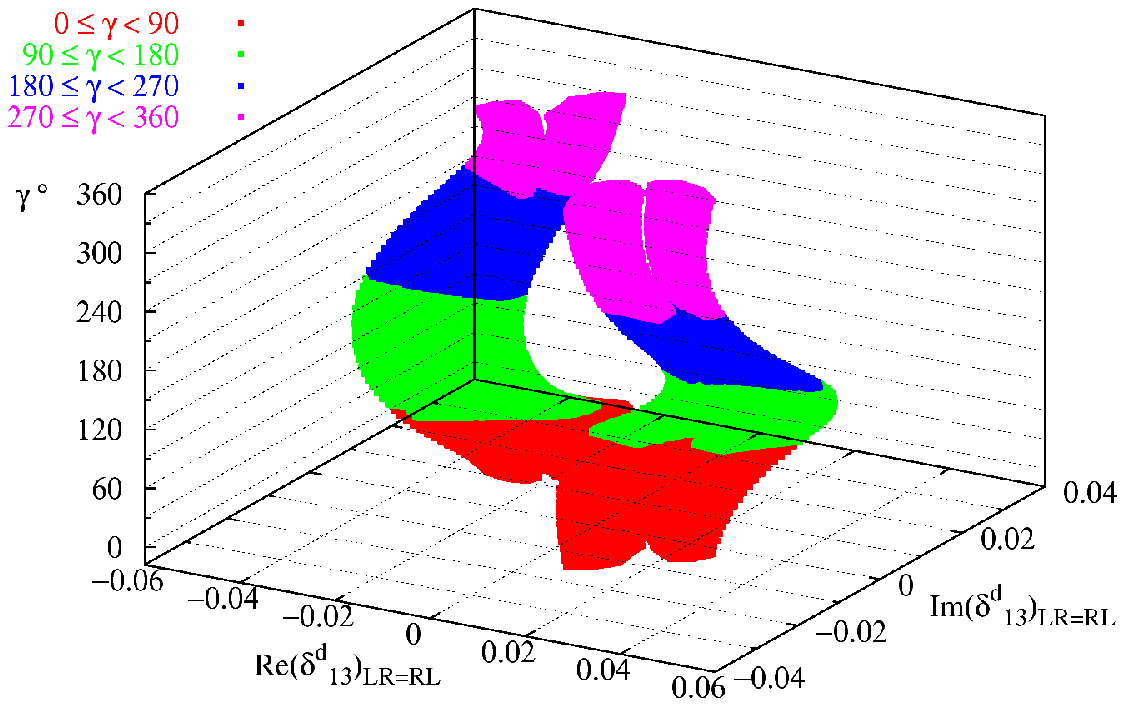}\\
\epsfxsize=0.45\textwidth\epsffile{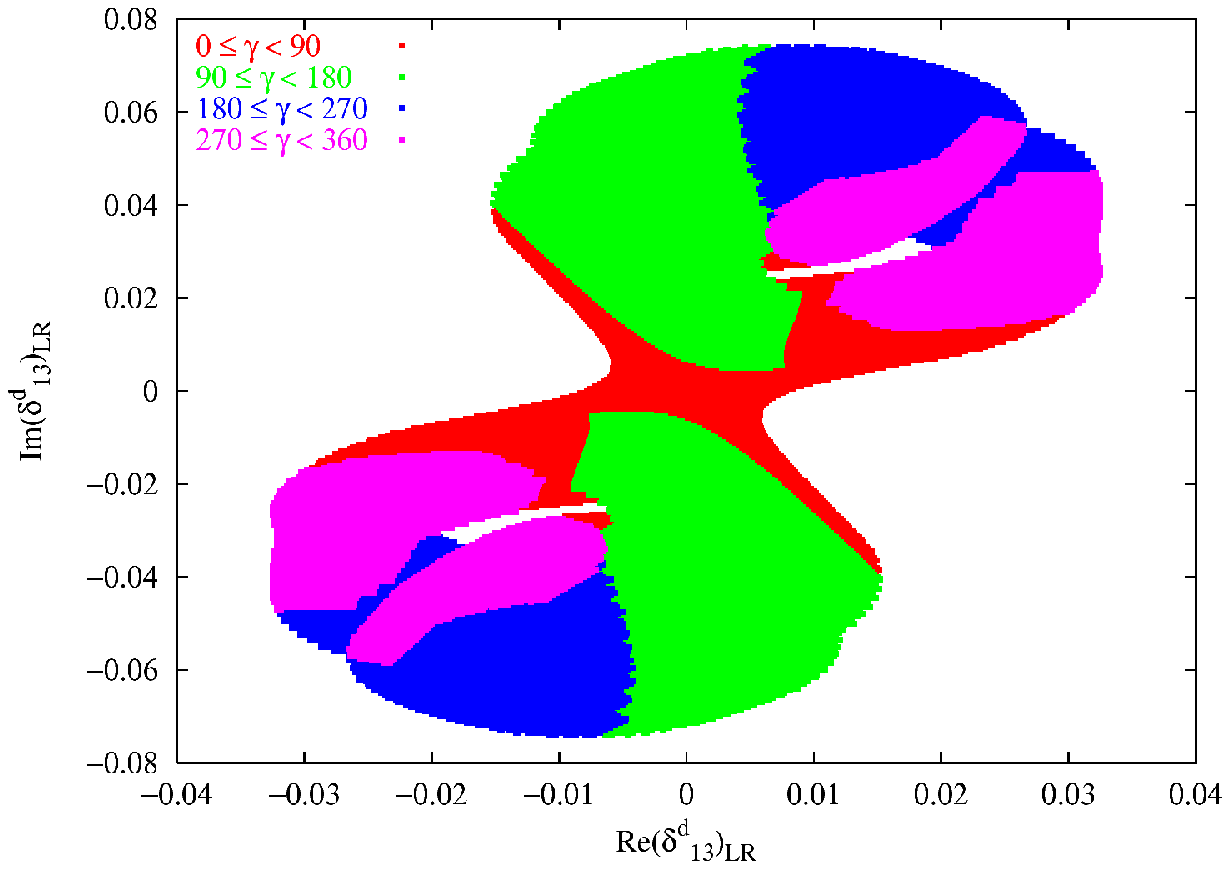} &\epsfxsize=0.45\textwidth \epsffile{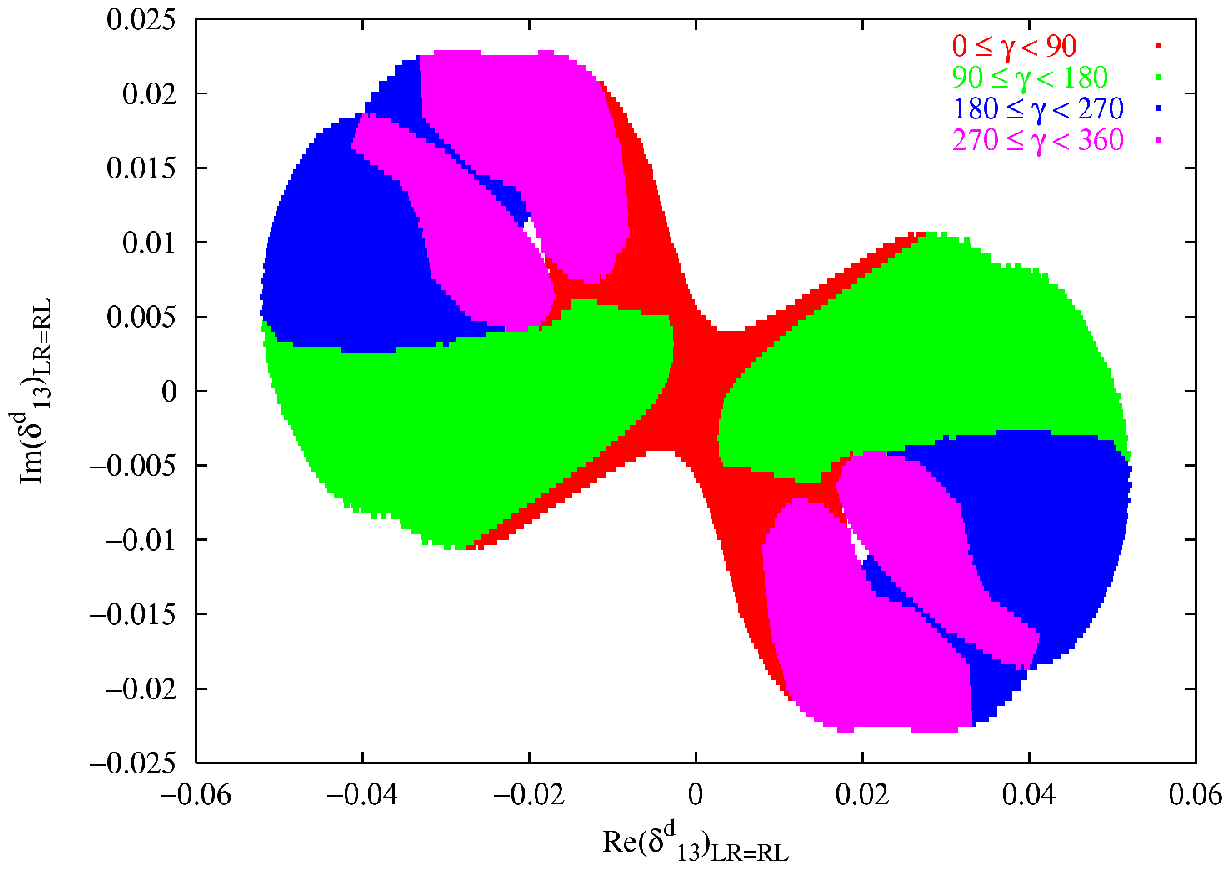}\\
\end{tabular}
\end{center}
\caption{{\sl Allowed regions in the ($\gamma$,  $\Re(\delta^{d}_{13})_{LR}$, $\Im(\delta^{d}_{13})_{LR}$) space
with $(\delta^{d}_{13})_{LR}$ only (left) and
$(\delta^{d}_{13})_{LR}=(\delta^{d}_{13})_{RL}$
(right). The two lower plots are the corresponding projections in the
$\Re(\delta^{d}_{13})_{LR}$--$\Im(\delta^{d}_{13})_{LR}$ plane. Different colours denote
 values of $\gamma$ belonging to different quadrants.
}}
\label{fig:LR}
\end{figure}

\section{Numerical Analysis of $\Delta B =2$ Processes}
\label{sec:numerics}
In this section, we illustrate the procedure followed in our analysis and give the main
results.
$\Delta m_d$ and $\sin 2\beta$ are defined in terms of the matrix element
$\langle\bar  B_d \vert{\cal H}_{\rm eff}^{\Delta B=2} \vert B_d \rangle$
which we schematically write as
\bea
 \label{schema}
 \langle\bar  B_d \vert{\cal H}_{\rm eff}^{\Delta B=2} \vert B_d \rangle&=&
 \Re{\cal A}_{SM} + i\, \Im{\cal A}_{SM}+\nn\\
 &&{\cal A}_{SUSY}\Re(\delta^d_{13})_{AB}^2+ i\, {\cal A}_{SUSY}\Im (\delta^d_{13})_{AB}^2\,,
\eea
where ${\cal A}_{SM}$  is the Standard Model contribution which depends on
the CKM matrix parameters, in particular on the CP violation phase $\gamma$.
$(\delta^d_{13})_{AB}$ denotes the generic effective coupling appearing in eq.~(\ref{inicoeff}).
In the following, for simplicity, the constraints are obtained by imposing that
the sum of the SUSY contributions proportional to a single $(\delta^d_{13})_{AB}$ parameter and
the SM contribution does not exceed by more than $1\sigma$ the experimental value
of  $\Delta m_d$  and $\sin 2\beta$.
This is justified (a posteriori) by noting that the constraints on
different $(\delta^d_{13})_{AB}$ parameters exhibit a hierarchical
structure, and therefore interference effects between different
contributions would require a large amount of fine tuning.
We consider the following four cases:
i)  $(\delta^{d}_{13})_{LL}$ only; ii) $(\delta^{d}_{13})_{LL}=(\delta^{d}_{13})_{RR}$;
iii) $(\delta^{d}_{13})_{LR}$ only; iv) $(\delta^{d}_{13})_{LR}=(\delta^{d}_{13})_{RL}$.
For this reason, in all the cases considered here, the amplitude in eq.~(\ref{schema}) only
depends on a single $(\delta^d_{13})_{AB}^2$.

\begin{table}[t]
\begin{center}
\begin{tabular}{||c|c|c|c|c|c|c||}  \hline \hline
& \multicolumn{3}{c|}{$\vert\Re(\delta^{d}_{13})_{LL}\vert$ }
& \multicolumn{3}{c|}{$\vert\Re(\delta^{d}_{13})_{LL=RR}\vert$}\\
\hline
$x$ & TREE & LO & NLO & TREE & LO & NLO \\

0.25& $4.9\times 10^{-2}$ & $5.4\times 10^{-2}$

& $6.2\times 10^{-2}$

& $3.1\times 10^{-2}$ & $2.0\times 10^{-2}$

& $1.9\times 10^{-2}$
\\
1.0& $1.1\times 10^{-1}$ & $1.2\times 10^{-1}$

& $1.4\times 10^{-1}$

& $3.4\times 10^{-2}$ & $2.2\times 10^{-2}$

& $2.1\times 10^{-2}$
\\
4.0& $6.0\times 10^{-1}$ & $6.7\times 10^{-1}$

& $7.0\times 10^{-1}$

& $4.7\times 10^{-2}$ & $3.0\times 10^{-2}$

& $2.8\times 10^{-2}$ \\
\hline
& \multicolumn{3}{c|}{$\vert\Im(\delta^{d}_{13})_{LL}\vert$ }
& \multicolumn{3}{c|}{$\vert\Im(\delta^{d}_{13})_{LL=RR}\vert$}\\
\hline

$x$ & TREE & LO & NLO & TREE & LO & NLO\\

0.25& $1.1\times 10^{-1}$ & $1.2\times 10^{-1}$

& $1.3\times 10^{-1}$

& $1.3\times 10^{-2}$ & $8.0\times 10^{-3}$

& $8.0\times 10^{-3}$
\\
1.0& $2.6\times 10^{-1}$ & $2.8\times 10^{-1}$

& $3.0\times 10^{-1}$

& $1.5\times 10^{-2}$ & $9.0\times 10^{-3}$

& $9.0\times 10^{-3}$
\\
4.0& $2.6\times 10^{-1}$ & $2.9\times 10^{-1}$

& $3.4\times 10^{-1}$

& $2.0\times 10^{-2}$ & $1.3\times 10^{-2}$

& $1.2\times 10^{-2}$
\\
\hline \hline
& \multicolumn{3}{c|}{$\vert\Re(\delta^{d}_{13})_{LR}\vert$ }
& \multicolumn{3}{c|}{$\vert\Re(\delta^{d}_{13})_{LR=RL}\vert$}\\
\hline

$x$ & TREE & LO & NLO & TREE & LO & NLO\\ 

0.25& $3.4\times 10^{-2}$ & $2.7\times 10^{-2}$

& $3.0\times 10^{-2}$ 

& $3.8\times 10^{-2}$ & $2.7\times 10^{-2}$

& $2.6\times 10^{-2}$ 
\\ 
1.0& $3.9\times 10^{-2}$ & $3.0\times 10^{-2}$

& $3.3\times 10^{-2}$

& $8.3\times 10^{-2}$ & $5.4\times 10^{-2}$

& $5.2\times 10^{-2}$ 
\\ 
4.0& $5.3\times 10^{-2}$ & $4.1\times 10^{-2}$

& $4.5\times 10^{-2}$ 

& $1.2\times 10^{-1}$ & $2.5\times 10^{-1}$

& $-$  \\
\hline
& \multicolumn{3}{c|}{$\vert\Im(\delta^{d}_{13})_{LR}\vert$ }
& \multicolumn{3}{c|}{$\vert\Im(\delta^{d}_{13})_{LR=RL}\vert$}\\
\hline

$x$ & TREE & LO & NLO & TREE & LO & NLO\\

0.25& $7.6\times 10^{-2}$ & $6.0\times 10^{-2}$

& $6.6\times 10^{-2}$

& $1.5\times 10^{-2}$ & $9.0\times 10^{-3}$

& $9.0\times 10^{-3}$
\\
1.0& $8.7\times 10^{-2}$ & $6.6\times 10^{-2}$

& $7.4\times 10^{-2}$

& $3.6\times 10^{-2}$ & $2.4\times 10^{-2}$

& $2.3\times 10^{-2}$
\\
4.0& $1.2\times 10^{-1}$ & $9.2\times 10^{-2}$

& $1.0\times 10^{-1}$

& $2.7\times 10^{-1}$ & $5.7\times 10^{-1}$

& $-$
\\
\hline  \hline
 \end{tabular}
\end{center}
 \caption{{\sl Maximum allowed values for   $\vert\Re\left(\delta^d_{ij}
 \right)_{AB}\vert$ and $\vert\Im\left(\delta^d_{ij}\right)_{AB}\vert$, with $A,B=(L,R)$, for an
  average squark mass $m_{\tilde{q}}=500$ GeV and for different values
 of $x=m_{\tilde{g}}^2/m_{\tilde{q}}^2$. We give the results  in the following
 cases:
 i) with the tree level Wilson coefficients, namely without
 evolution from $M_{S}$ to $m_{b}$, and VIA $B$ parameters, denoted by
 TREE;
 ii) with LO evolution and VIA $B$ parameters, denoted by LO;
 iii) with NLO evolution and lattice $B$ parameters, denoted by NLO. The
missing entries correspond to cases in which no constraint was found for
$\vert\left(\delta^d_{ij}\right)_{AB}\vert < 0.9$.} }
\label{tab:reds2_500} \end{table}

\begin{table}[t]
\begin{center}
 \begin{tabular}{||c|c|c|c|c||}  \hline \hline
& \multicolumn{2}{c|}{$\vert\Re(\delta^{d}_{13})_{LL}\vert$ }
& \multicolumn{2}{c|}{$\vert\Re(\delta^{d}_{13})_{LL=RR}\vert$}\\
\hline

$x$ & $m_{\tilde q}=250$ GeV & $m_{\tilde q}=1000$ GeV
& $m_{\tilde q}=250$ GeV & $m_{\tilde q}=1000$ GeV \\

0.25& $-$

& $1.3\times 10^{-1}$ 

& $-$ 

& $3.8\times 10^{-2}$ 
\\ 
1.0& $6.5\times 10^{-2}$ 

& $3.1\times 10^{-1}$ 

& $1.0\times 10^{-2}$ 

& $4.2\times 10^{-2}$ 
\\ 
4.0& $3.2\times 10^{-1}$ 

& $-$

& $1.4\times 10^{-2}$ 

& $5.9\times 10^{-2}$ \\
\hline 
& \multicolumn{2}{c|}{$\vert\Im(\delta^{d}_{13})_{LL}\vert$ }
& \multicolumn{2}{c|}{$\vert\Im(\delta^{d}_{13})_{LL=RR}\vert$}\\ 
\hline 

$x$ & $m_{\tilde q}=250$ GeV & $m_{\tilde q}=1000$ GeV
& $m_{\tilde q}=250$ GeV & $m_{\tilde q}=1000$ GeV \\ 

0.25& $-$ 

& $2.8\times 10^{-1}$ 

& $-$

& $1.6\times 10^{-2}$ 
\\ 
1.0& $1.3\times 10^{-1}$ 

& $5.0\times 10^{-1}$ 

& $4.0\times 10^{-3}$ 

& $1.8\times 10^{-2}$ 
\\ 
4.0& $1.5\times 10^{-1}$ 

& $-$ 

& $6.0\times 10^{-3}$ 

& $2.5\times 10^{-2}$ 
\\ 
\hline \hline 
& \multicolumn{2}{c|}{$\vert\Re(\delta^{d}_{13})_{LR}\vert$ }
& \multicolumn{2}{c|}{$\vert\Re(\delta^{d}_{13})_{LR=RL}\vert$}\\ 
\hline 

$x$ & $m_{\tilde q}=250$ GeV & $m_{\tilde q}=1000$ GeV
& $m_{\tilde q}=250$ GeV & $m_{\tilde q}=1000$ GeV \\ 

0.25& $-$ 

& $6.2\times 10^{-2}$ 

& $-$

& $5.3\times 10^{-2}$ 
\\ 
1.0& $1.5\times 10^{-2}$ 

& $6.8\times 10^{-2}$ 

& $2.5\times 10^{-2}$ 

& $1.1\times 10^{-1}$ 
\\ 
4.0& $2.2\times 10^{-2}$ 

& $9.4\times 10^{-2}$ 

& $-$ 

& $-$  \\
\hline 
& \multicolumn{2}{c|}{$\vert\Im(\delta^{d}_{13})_{LR}\vert$ }
& \multicolumn{2}{c|}{$\vert\Im(\delta^{d}_{13})_{LR=RL}\vert$}\\ 
\hline 

$x$ & $m_{\tilde q}=250$ GeV & $m_{\tilde q}=1000$ GeV
& $m_{\tilde q}=250$ GeV & $m_{\tilde q}=1000$ GeV \\ 

0.25& $-$ 

& $1.4\times 10^{-1}$ 

& $-$ 

& $2.1\times 10^{-2}$ 
\\
1.0& $3.3\times 10^{-2}$ 

& $1.5\times 10^{-1}$ 

& $9.0\times 10^{-3}$ 

& $4.5\times 10^{-2}$ 
\\ 
4.0& $4.8\times 10^{-2}$ 

& $2.2\times 10^{-1}$ 

& $-$ 

& $-$ 
\\
\hline  \hline
 \end{tabular}
\end{center}
\caption{{\sl Maximum allowed values for   $\vert\Re\left(\delta^d_{ij}
\right)_{AB}\vert$ and $\vert\Im\left(\delta^d_{ij}\right)_{AB}\vert$, with $A,B=(L,R)$, for an
average squark mass $m_{\tilde{q}}=250$~GeV or $1000$~GeV and for different values
of $x=m_{\tilde{g}}^2/m_{\tilde{q}}^2$. We give the results
with NLO evolution and lattice $B$ parameters.  The
missing entries correspond to cases in which no constraint was found for
$\vert\left(\delta^d_{ij}\right)_{AB}\vert < 0.9$.} }
\label{tab:reds2_1250}
\end{table}

The procedure used in our numerical analysis is the following:
\begin{itemize}
\item We scan over all  possible values of $\gamma$ between $0$ and $2\pi$
since, in the presence of SUSY effects, this parameter is no longer constrained by
the UT analysis. Indeed, $\epsilon_{K}$ may be subject to large SUSY contributions due to the 
$(\delta^{d}_{12})_{AB}$ couplings and the value of $\gamma$ can be very different from the SM value;
\item for a fixed value of $\gamma$, we scan over $\Re(\delta^{d}_{13})_{AB}$ and
$\Im(\delta^{d}_{13})_{AB}$ between $-1$ and~$1$;
\item we let the other experimental and theoretical parameters, in particular the $B$ parameters,
  vary within the ranges given in Table~\ref{tab:parameters} and in Eq.~(\ref{eq:fres});
\item in the ($\gamma$,  $\Re(\delta^{d}_{13})_{AB}$, $\Im(\delta^{d}_{13})_{AB}$)
space, we select the region where the predicted values of  $\Delta m_d$  and $\sin 2\beta$
lie within $1\sigma$ from the measured values.
\end{itemize}
As can be seen from Eq.~(\ref{schema}), the amplitude only depends on 
the square of the   coupling,   $(\delta^{d}_{13})_{AB}^2$.
Therefore, the selected values of $(\delta^{d}_{13})_{AB}$ have a twofold ambiguity
($\Re(\delta^{d}_{13})_{AB}\to -\Re(\delta^{d}_{13})_{AB}$ and
 $\Im(\delta^{d}_{13})_{AB}\to -\Im(\delta^{d}_{13})_{AB}$)
as will be evident from our results.

\begin{figure}[t]
\begin{center}
\begin{tabular}{c c}
\epsfxsize=0.45\textwidth\epsffile{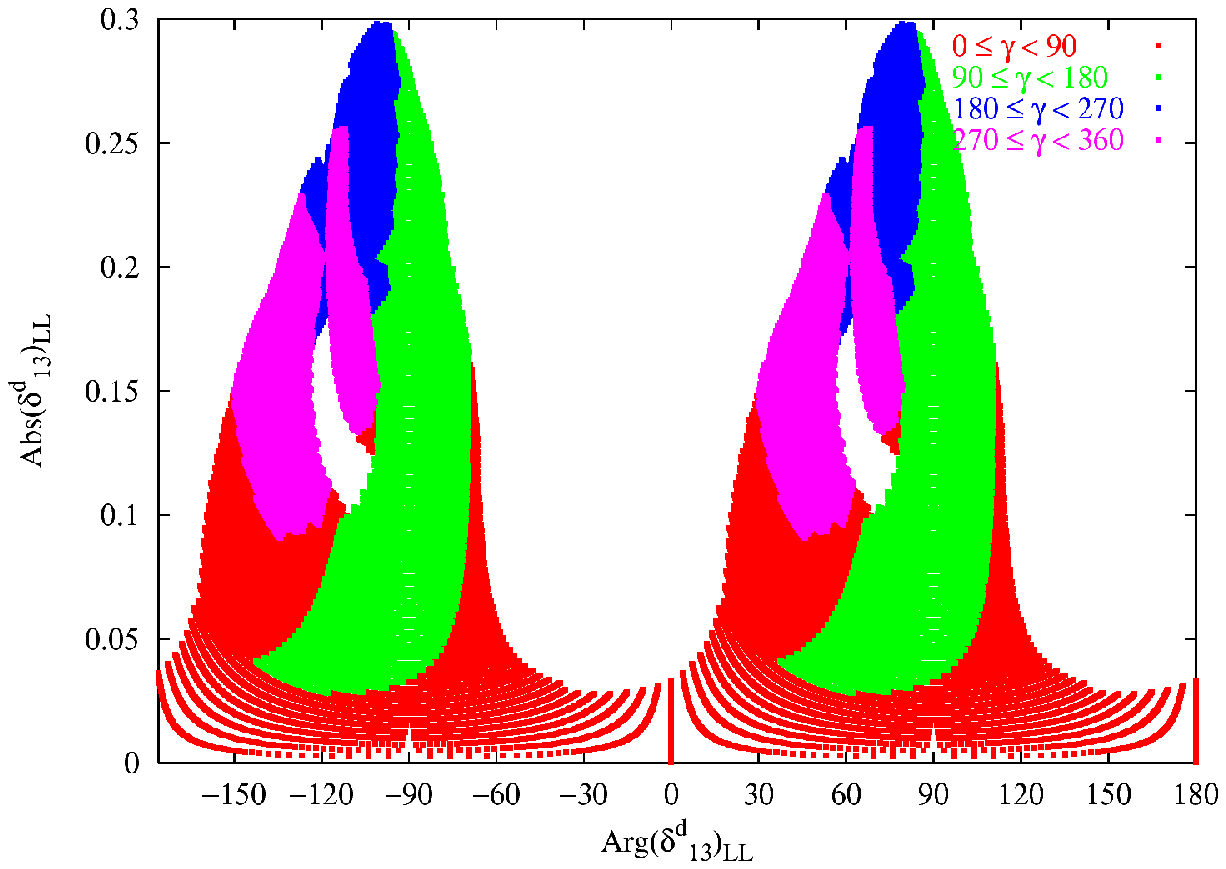} & \epsfxsize=0.45\textwidth\epsffile{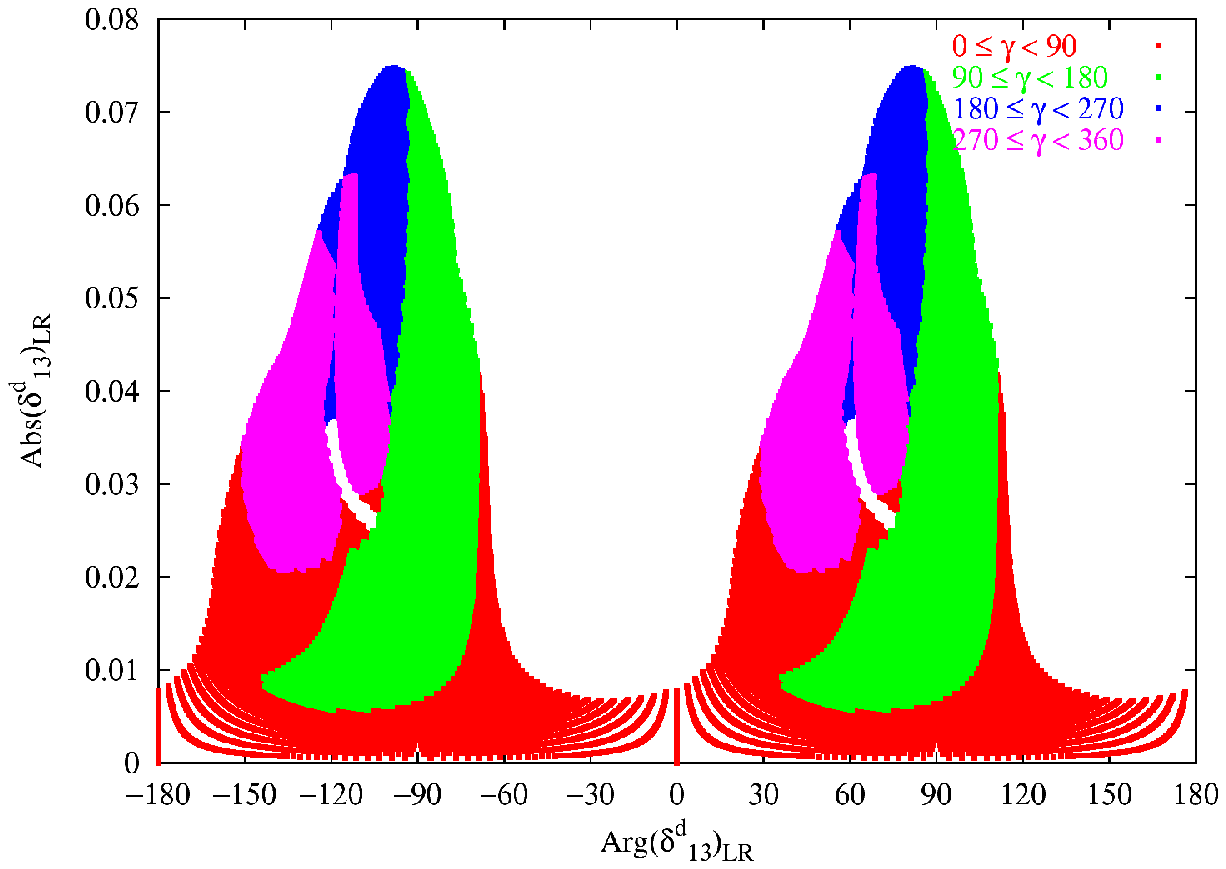}\\
\epsfxsize=0.45\textwidth\epsffile{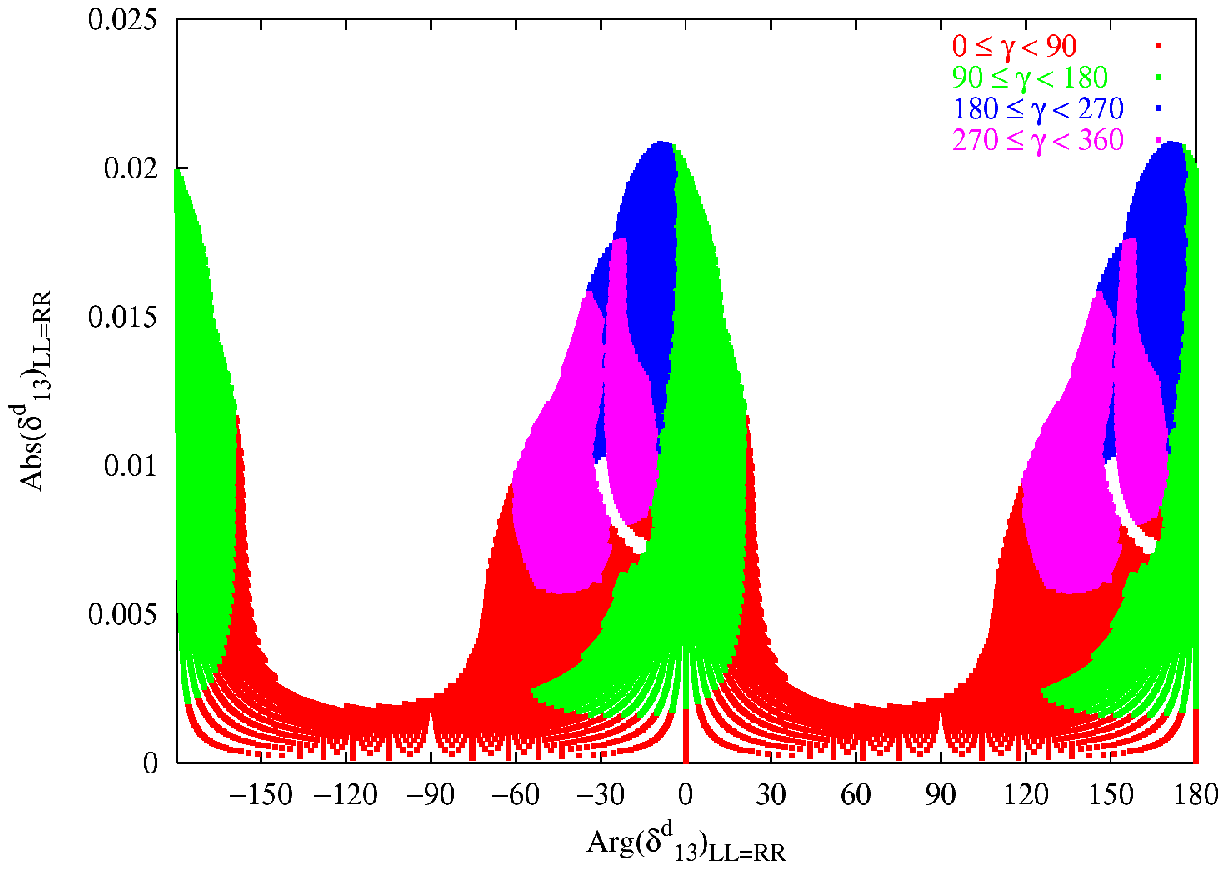} &\epsfxsize=0.45\textwidth \epsffile{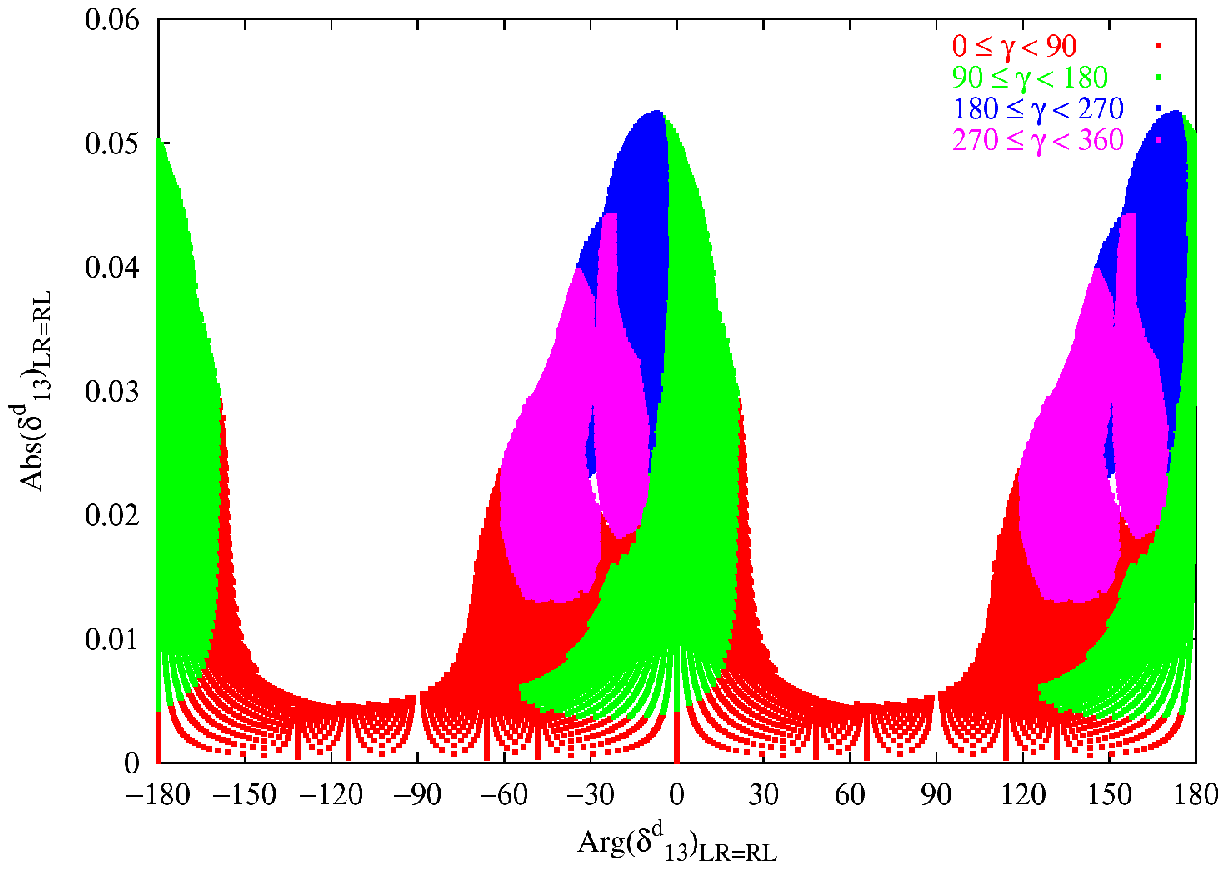}\\
\end{tabular}
\end{center}
\caption{{\sl Allowed values of Abs$(\delta^{d}_{13})_{LR}$ as a function of Arg$(\delta^{d}_{13})_{AB}$
in the four considered cases.  Different colours denote values of $\gamma$ belonging to different quadrants.
}}
\label{fig:abs}
\end{figure}

We are ready to discuss the physics results. In 
Figs.~\ref{fig:LL}--\ref{fig:LR}, for $m_{\tilde q}=m_{\tilde g}=500$~GeV, we show the allowed regions
in the  ($\gamma$,  $\Re(\delta^{d}_{13})_{AB}$, $\Im(\delta^{d}_{13})_{AB}$) space and
the corresponding projections on the  ($\Re(\delta^{d}_{13})_{AB}$, $\Im(\delta^{d}_{13})_{AB}$)
plane. For the readers' convenience, we have coloured regions corresponding 
to $\gamma$ in the four
quadrants with different colours.
An alternative way of presenting our results is given in Fig.~\ref{fig:abs} where
the allowed values of Abs$(\delta^{d}_{13})_{AB}$ are plotted as a function of  Arg$(\delta^{d}_{13})_{AB}$ .

We give the
maximum allowed values of $\vert\Re(\delta^{d}_{13})_{AB}\vert$ and
$\vert\Im(\delta^{d}_{13})_{AB}\vert$
in Table~\ref{tab:reds2_500} for $x=0.25$, $1$ and $4$.
We also give results for $m_{\tilde q}=250$~GeV and
$1000$~GeV in Table~\ref{tab:reds2_1250}.
We find that the constraints on the $(\delta^{d}_{13})_{AB}$ parameters are less
effective by a factor from 10 to 100 with respect to those extracted on $(\delta^{d}_{12})_{AB}$ from $\Delta S=2$
transitions. The reason for this difference is twofold. On the one hand, one naturally expects constraints of the order of
$\sin^{3}\theta_c$ for $B_d$ -- $\bar B_d$ mixing and  $\sin^5\theta_c$ for $K^0$--$\bar K^0$ mixing,
dictated by the top contribution in the SM amplitude. On the other, in the  $K^0$--$\bar K^0$ case, LR contributions
are chirally enhanced, so that limits coming from chirality-changing operators are more effective.
Obviously, specific models of flavour, where $(\delta^{d}_{12})_{AB}$ and $(\delta^{d}_{13})_{AB}$  are
correlated, can constrain the squark-mass splittings more severely. In these cases one could even attempt the
inclusion of chargino contributions to the mixing~\cite{Chua:2001dd,Masiero:2001cc}. Our general formulae
for the effective Hamiltonian and the magic numbers remain valid, and the phenomenological analysis can be
carried out along the same lines.

We now discuss the difference between our results and those obtained with tree level or LO Wilson coefficients 
and VIA matrix elements. In this case, with our definition of the $B$ parameters
given in eq.~(\ref{eq:bpars}), with $\kappa= (m_{b}/m_{B_{d}})^{2}=0.76$, 
we have used
\be
\label{eq:fresvia}
\begin{tabular}{ll}
$B_{1}^{VIA} = 1.0$\, ,   & $B_{2}^{VIA} = 1.0$\, ,  \\
$B_{3}^{VIA} = 1.0$\, ,  &  $B_{4}^{VIA} = 1.13$\, , \\
$B_{5}^{VIA} = 2.14$\, .  & \\
\end{tabular}
\ee

With NLO corrections and lattice $B$ parameters, the constraints
are generically looser than using LO and VIA $B$ parameters.
The theoretical uncertainties of the NLO constraints are, however,
much smaller. In the LL=RR case, the NLO bounds are even more stringent
than the LO ones, in spite of the uncertainties on the $B$ parameters.
The most striking effect is obtained in the LR=RL case for $x=4$.
Here, the strong cancellation taking place at the LO is exacerbated by
NLO corrections, jeopardising any possibility to constrain $(\delta^{d}_{13})_{LR}$
in this case. Clearly, given the cancellation, this result should be interpreted
with care and verified by calculations in any given model.

With the future measurements of $\Delta m_{s}$ at  proton
colliders,  and hopefully of the corresponding CP violating effects
in $B_{s}$ mixing and decays, it will be soon possible to perform a
similar analysis for the $B_{s}$ system.  Work along this line is in
progress~\cite{noifuture}.

\section{Conclusions}
\label{sec:concl}

In this work we have provided an improved determination of the
gluino-mediated SUSY contributions to $B_d$ -- $\bar B_d$ mixing and
to the CP asymmetry $a_{J/\psi K_s}$ in
the framework of the mass-insertion method. The improvement consists
in introducing the NLO QCD corrections to  ${\cal H}_{\rm
  eff}^{\Delta B=2}$~\cite{Ciuchini,Urban} and in replacing the VIA
matrix elements with their recent lattice computation~\cite{Damir}. As a
glimpse at Table~\ref{tab:reds2_500} readily
reveals, these improvements affect previous results in a different
way, according to the operators of ${\cal H}_{\rm eff}^{\Delta B=2}$
that one considers. The effect is particularly large for Left-Right
operators.

We have provided an analytic formula for the most general low-energy
${\cal H}_{\rm eff}^{\Delta B=2}$ at the NLO, in terms of the Wilson
coefficients at the high energy scale.
This formula can be readily
used to compute $\Delta m_d$ and $\sin 2\beta$ in any extension of
the SM with new heavy particles.

FCNC and CP violating phenomena (in particular in $B$ physics) are
promising candidates for some indirect SUSY signal before LHC, and are
in many ways complementary to direct SUSY searches. From this point of
view the theoretical effort to improve as much as possible our
precision on FCNC computations in a SUSY model-independent framework
is certainly worth and, hopefully, rewarding.

\section*{Acknowledgements}
We thank Fabrizio Parodi for information about the world average of
the direct measurements of $\sin 2 \beta$ and for discussions.
This work has been partially supported by the
European Community's Human potential programme ``Hadron Phenomenology from Lattice QCD"
contract number HPRN-CT-2000-00145 and by the
RTN European project ``Across the Energy Frontier" contract number
HPRN-CT-2000-0148.                                                              


\end{document}